\documentclass[sigconf]{acmart}

\AtBeginDocument{%
  }

\setcopyright{acmcopyright}
\copyrightyear{2018}
\acmYear{2018}
\acmDOI{XXXXXXX.XXXXXXX}

\acmConference[ESEM '22]{Empirical Software Engineering journal}{Sun 18 - Fri 23}{Helsinki, Finland}
\acmPrice{15.00}
\acmISBN{978-1-4503-XXXX-X/18/06}

\usepackage{siunitx}
\usepackage{enumitem}
\usepackage{amsmath,amsfonts}
\usepackage{algorithmic}
\usepackage{graphicx}
\usepackage{textcomp}
\usepackage{xcolor}
\usepackage{booktabs}
\usepackage{hyperref}

\newcommand{\rev}[1]{\textcolor{black}{#1}}

\begin{document}

\title{Studying the explanations for the automated prediction of bug and non-bug issues using LIME and SHAP}

\author{Benjamin Ledel}
\email{benjamin.ledel@tu-clausthal.de}
\affiliation{%
  \institution{TU Clausthal}
  \streetaddress{Arnold-Sommerfeld-Straße 1}
  \city{Clausthal-Zellerfeld}
  \state{Lower Saxony}
  \country{Germany}
  \postcode{38678}
}

\author{Steffen Herbold}
\email{steffen.herbold@tu-clausthal.de}
\affiliation{%
  \institution{TU Clausthal}
  \streetaddress{Arnold-Sommerfeld-Straße 1}
  \city{Clausthal-Zellerfeld}
  \state{Lower Saxony}
  \country{Germany}
  \postcode{38678}
}


\begin{abstract}

  \noindent\textbf{Context:} The identification of bugs within the reported issues in an issue tracker is crucial for the triage of issues. Machine learning models have shown promising results regarding the performance of automated issue type prediction. However, we have only limited knowledge beyond our assumptions how such models identify bugs. LIME and SHAP are popular technique to explain the predictions of classifiers. 

  \noindent\textbf{Objective:} We want to understand if machine learning models provide explanations for the classification that are reasonable to us as humans and align with our assumptions of what the models should learn. We also want to know if the prediction quality is correlated with the quality of explanations. 

  \noindent \textbf{Method:} We conduct a study where we rate LIME and SHAP explanations based on their quality of explaining the outcome of an issue type prediction model. For this, we rate the quality of the explanations themselves, i.e., if they align with our expectations and if they help us to understand the underlying machine learning model. 
\end{abstract}

\ccsdesc[500]{Computing methodologies~Reasoning about belief and knowledge}
\ccsdesc[300]{Computing methodologies~Causal reasoning and diagnostics}
\ccsdesc[100]{Software and its engineering~Software creation and management}

\maketitle

\section{Introduction}
While machine learning techniques are widely used, the interpretability and explainability of the resulting models are often limited \cite{guidotti2018survey} \cite{biran2017explanation}. Understanding the reason behind a decision or prediction of a model leads to trust and helps to validate the correctness and robustness of the model \cite{ribeiro2016should}. Ribeiro et al. \cite{ribeiro2016should} proposed \textit{Local Interpretable Model-agnostic Explanations} (LIME), a technique that provides an explanation for the predictions of any classification model. LIME has support for textual data and has a graphical representation of the generated explanations. It was validated and successfully used in several user studies to help the human to understand the machine learning models \cite{magesh2020explainable} \cite{goyal2017making} \cite{ribeiro2016model}. Lundberg et al. \cite{lundberg2017unified} proposed a unified approach to interpreting model using \textit{SHapley Additive exPlanations} (SHAP). SHAP supports text classification with deep neural networks. Kokalj et al. \cite{kokalj2021bert} outlined that LIME and SHAP are the most widely used permutation-based explanation methods.

Within the domain of software engineering, the identification of bugs among the issues reported in an issue tracking system is an important task to support the triage of issues, e.g., for the localization of bugs~\cite{mills2018bug} that can be supported by text mining. In this work, we will follow the guideline by Herzig et al.~\cite{herzig2013s} to decide if an issue is a bug or a non-bug. They defined a bug as an issue report that documenting corrective maintenance tasks that require semantic changes to the source code \cite{herzig2013s}. Using this characterization of bug issues, Herzig et al.~\cite{herzig2013s} provided clear guidelines, which are used to create ground truth data for issue types~\cite{herzig2013s, herbold2019issues} by researchers. Such data can be used to train and evaluated machine learning models, which provide promising results for the automated identification of bug issues~(e.g., \cite{herbold2020feasibility}, \cite{palacio2019learning}, \cite{von2021validity}). However, while the past research showed that machine learning models can achieve satisfactory performance, we only have a limited understanding in how the machine learning achieve this performance. 

Within this study, we will close this gap in our knowledge. We will validate the prediction of bug issues on a large scale and at the same time deepen our understanding of LIME and SHAP as tools for explaining machine learning models within a confirmatory study, based on our assumptions we derive from prior work on issue type prediction. We examine if the correctness of the predictions is related to explanation quality as expected, if it is easier to explain why an issue is a bug, than why it is not a bug, as well the general quality of the explanations. We evaluate the explanations of LIME and SHAP in terms of four qualitative categories, i.e., if the explanation is \textit{related} to the prediction, the terms used to explain are \textit{unambiguous}, the explanation captures the \textit{context}, and if we gain \textit{insights} into the machine learning algorithm.

The contributions of our planned research project are the following:

\begin{itemize}
  \item A large data set of rated LIME and SHAP explanations of issue type predictions based on the quality of their explanations. The data will contain 3.092 issues with the prediction whether they are a bug or not from the machine learning models and their corresponding LIME and SHAP explanation. 
  \item A confirmatory analysis if the LIME and SHAP explanations for the identification of bug issues meet the expectations that we have on how such machine learning models, as well as LIME and SHAP, to work. 
  \item Insights into the quality of issue prediction models beyond the prediction performance, that help us to better understand how such models work and, thereby, better understand in which context such models can be successfully applied. 
\end{itemize}

The remainder of this paper is structured as follows. In Section \ref{sec:relatedWork}, we give a short overview of the related work. Next, we introduce our research questions and hypotheses in Section \ref{sec:research-questions}. Then, we describe our research protocol, including materials, variables, execution plan, and analysis plan in Section \ref{sec:research-protocol}. In Section \ref{sec:limitation}, we describe the limitations and conclude the paper with an overview of the generated data.

\section{Related work}
\label{sec:relatedWork}
Explainable methods can be divided into global and local explainability \cite{guidotti2018survey}. The goal of the global techniques is to provide an understanding of the whole model. The explanation is able to explain all entries from the data set and can also applied to generic new data. Local explanation, instead, have the goal to explain only a single instance or small amount of entries from the data set \cite{visani2020statistical}.   

In our study, we consider explanations that are generated using LIME and SHAP. Ribeiro et al. \cite{ribeiro2016should} proposed LIME as a technique to explain the predictions of any classifier in an interpretable and faithful manner. Thereby, LIME perturbs the input around the local prediction and generates the neighborhood of instances. These generated instances are weighted according the proximity to the original instance. Finally, a linear model that approximates the model well for the original instances is calculated.  The technique was already used in different domains, for example music content analysis \cite{mishra2017local} or to explain classification of lymph node metastases \cite{palatnik2019local}. Furthermore, LIME has promising results in the domain of defect prediction \cite{jiarpakdee2021practitioners}. Thus, in this study, we will extend this scope to the domain of using LIME with issues from software projects, especially for the classification of issue types in bugs and non-bugs. Beside LIME, we will use SHAP as second explanation algorithm. Lundberg et al. \cite{lundberg2017unified} proposed SHAP as a unified method to explain predictions of different machine learning models. SHAP uses concepts from cooperative game theory and is able to assign each attribute from the input of a machine learning model an importance value based on its impact on the prediction when the feature is present or not during the SHAP estimation \cite{lundberg2017unified}. In order to explain different type of models, SHAP has different variants like Linear SHAP, Low-Order SHAP, Max SHAP or Deep SHAP. These variants help to reduce the compute time to calculate the model-specific explanation. In this work, we will use the variant Deep SHAP for natural language models.   
Lundberg et al. \cite{lundberg2017unified} noted that LIME is a special case of SHAP.     

Tantithamthavorn et al. \cite{tantithamthavorn2020explainable} outlined that explainable AI is very important for software engineering. The main purpose in the current literature is in the area of defective prediction. Multiple studies have shown that LIME was able to help developers to localize which lines of code are the most risky and explain the prediction of defect models \cite{tantithamthavorn2020explainable, jiarpakdee2021practitioners}. However, these studies worked with tabular data, in comparison to our application to text mining.  Extending the idea of LIME, Pornprasit et al. \cite{pornprasit2021pyexplainer} proposed PyExplainer, a local rule-based model-agnostic technique for generating explanation of Just-In-Time defect predictions. In two case studies PyExplainer outperforms LIME in explaining the local instance by improving for example the synthetic neighbor generation. However, PyExplainer \rev{works only with tabular data and} has no support for textual data.  

An important task in the training of defect prediction models is the correct issue type classification. The potential large impact of the correct classifications on defect prediction research was first shown by Herzig et al. \cite{herzig2013s}. Herbold et al. \cite{herbold2020feasibility} summarized that knowledge about how issue type predictions work is still limited. To the best of our knowledge, no explainable AI techniques (including LIME), \rev{were} investigated for issue type predictions so far. 

\section{Research questions and hypotheses}
\label{sec:research-questions}
Considering our primary goal of the study, we state the following research question and hypotheses.
\begin{itemize}[leftmargin=*]
\item[] \textbf{RQ:} How is the correctness of the prediction and the quality of the explanation correlated? 
 \begin{itemize}
        \item[] \textbf{H1} Correct predictions of bugs have a higher qualitative score than correct predictions of non-bugs.
        \item[] \textbf{H2} The projects have no direct influence on the qualitative scores. 
        \item[] \textbf{H3} The qualitative scores of SHAP are higher than the scores of LIME
      \end{itemize}
\end{itemize}
We derived Hypothesis \textbf{H1} from the definition of bugs within the guidelines by Herzig et al.~\cite{herzig2013s}, which define how to identify bugs within issue tracking systems. While the criteria for bugs are relatively focused (e.g., dereferencing a null pointer, memory issues, or other crashes against the criteria), there are many different types of issues for non-bugs, e.g., changing requirements, requests for new features, suggestions of refactorings, architectural considerations, documentation tasks, licensing issues, and so forth. Consequently, when a machine learning model needs to decide if an issue is a bug or not, we believe that it focuses on identifying characteristics of bugs (e.g., occurrence of exceptions, dissatisfaction of users, mention of mistakes). We hypothesize that if issues exhibit the characteristics of bugs, they are classified as such, otherwise, they are classified as non-bugs. If we are correct, this would mean that the explanations for bug predictions should focus on the above mentioned characteristics on bugs and be helpful. However, it also follows that predictions of non-bugs are the result of the absence of a signal, which means that LIME and SHAP would not be able to find a good explanation, as there is no signal to explain. 

We derived Hypothesis \textbf{H2} from the work by Ribeiro et al. \cite{ribeiro2016model}. They demonstrated that LIME is able to provide insights to the user of a machine learning model for different domains. They show that it helps the user to understand the model and its behavior to build trust \cite{ribeiro2016model}. However, Alvarez-Melis et al. \cite{alvarez2018robustness} showed that LIME is not always stable and Zhang et al. \cite{zhang2019should} \rev{investigated potential reasons for the lack of robustness.} They showed that a potential source is the randomness in its sampling procedure. Overall, we assume that this source has smaller effect on the quality of the explanation than the other sources like the machine learning model itself. Thus, the qualitative score should be stable across different projects for the same machine learning model. 
For SHAP explanations, Yuan et al. \cite{man2021best} showed that the stability of SHAP lies on the background sample size. With an increase of the background sample size the stability increases. Since, we plan to investigate a large data set and are not aware of any further restrictions that the arguments of LIME \rev{are} not applicable for SHAP, we derived that the Hypothesis also holds for SHAP. 

We derived Hypothesis \textbf{H3} from the work Kokalj et al. \cite{kokalj2021bert} and Lundberg et al. \cite{lundberg2017unified}. Lundberg et al. \cite{lundberg2017unified} derived the explanation algorithm SHAP as a general variant of LIME, which is suitable and tested for deep neuronal networks. Therefore, we believe that SHAP outperforms LIME. However, Riberiro et al.\cite{ribeiro2016model} proposed LIME as an explanation tool for any classifier. Lundberg et al. \cite{lundberg2017unified} compared explanations from LIME and SHAP to validate the consistency with the human intuition. They found a much stronger agreement between human explanations and SHAP than with other methods \cite{lundberg2017unified}. However, we are not aware of an independent confirmation of this findings for the domain of issue type prediction. 
Slack et al. \cite{slack2020fooling} outlined that LIME is more vulnerable against fools than SHAP. Fools are  constructed adversarial classifiers that can fool post hoc explanation techniques which rely on input perturbations \cite{slack2020fooling}. Thus, the stability of LIME on a large dataset could suffer from the vulnerabilities, which leads to a inferior performance.   

\section{Research protocol}

We now define the materials, variables, execution plan, and analysis plan of our research protocol. 

\label{sec:research-protocol}
\subsection{Materials}
In the following section, we describe the materials including the data set of issues and their issue type predictions \rev{of the machine learning model}. Next, we describe the subjects, namely the LIME and SHAP explanations of issue type predictions.

\subsubsection{Predictions of bug and non-bug issues}
\label{sec:models}
\rev{Herbold et al. \cite{herbold2020feasibility} provide a data set of 30,922 issues with 11,154 manually validated bugs from 38 open source projects.} For the prediction of issue types, we will use a fine-tuned seBERT model, which was demonstrated by Trautsch et al. \cite{trautsch2022predicting} to outperform the best performing approach from a recent benchmark \cite{herbold2020feasibility}. Since we only want to have an output of bug or non-bug, we will fine-tune the model with a training set from the data set with manually validated bugs. We will use the data set from Herbold et al. \cite{herbold2020feasibility}, which is to the best of our knowledge, the largest data set with manually validated bugs.

\subsubsection{Subjects}
\label{sec:limeexplanations}
The subjects of this study are the LIME and SHAP explanations of issue type predictions. We plan to investigate the explanatory power of LIME and SHAP to examine the relationship between the prediction performance of the model and the quality of the LIME and SHAP explanation. Furthermore, we plan to compare the quality of the LIME and the SHAP explanations. We use stratified sampling of issues for each project, since the class of bug issues is smaller than the class of non-bugs \cite{taherdoost2016sampling}. We use 10\% of the issues from each project in our study to ensure that the sample contains issues from each project, i.e., 3092 issues. Thus, we calculate and rate $2 \cdot 3,092 = 6,184$ LIME and SHAP explanations in our study. We estimate the effort to rate the explanation of a single issue between 5 and 10 minutes. Based on this assumption, the total amount of work for a single rater is at least $5~\frac{\text{minutes}}{\text{issue}} \cdot 3,092~\text{issues} = 15,460~\text{minutes} = 257,7~\text{hours} \approx 33~\text{days}.$ \rev{The maximum of work for a single rater is $10~\frac{\text{minutes}}{\text{issue}} \cdot 3,092~\text{issues} = 30,920~\text{minutes} = 515,4~\text{hours} \approx 66~\text{days}.$ The time estimation per issue can vary significantly based on the clarity, length and detail level in the issue.}

Prior studies of LIME and SHAP explanations are limited to the amount of analysed explanation and limited in scope of different application, e.g. highlighting lines of defect predictions to support the developer \cite{wattanakriengkrai2020predicting}. We broaden this scope in our study to the task of issue type classification and in the number of rated explanations. Further restrictions are that there are only few studies who compared LIME and SHAP explanations against each other in a user study.  

\subsection{Variables}
\label{sec:participantstags}

We rate the LIME and SHAP explanations of predictions of the machine learning models based on a qualitative rating for each explanation for the four categories listed in Table~\ref{tab:tags}. For each category, we decide if the category applies (+1), is neutral (0), or does not apply (-1). The average qualitative rating is then computed by averaging over the four categories. We note that this rating is similar to a Likert scale~\cite{likert1932technique}. 

\begin{table*}[t]
\centering
\begin{tabular}{@{}lp{15cm}@{}}
\toprule
\textbf{Category}& \textbf{Explanation} \\ \midrule
Related & There is a clear relationship between the important words of explanation and the prediction of the algorithm. This relationship should follow the definition of a bug issue by Herzig et al.~\cite{herzig2013s}.\\
Unambigiuous & The explanation is unambiguous, if there are no words of mixed meaning or if all words are used in their correct meaning with respect to the explanation.  \\
Contextual & The important words of the explanation have a contextual relationship with each other. \\
Insightful  & The explanation offers insights into how the machine learning model works. \\ \midrule
\end{tabular}
\caption{List of categories for which a positive, neutral, or negative rating is performed for each explanation.}
\label{tab:tags}
\end{table*}
For each LIME and SHAP explanation, we will measure the following dependent variable with respect to the issue type prediction of the approaches.
\begin{itemize}
  \item $qs \in [-4, 4]$: The qualitative rating of the LIME and SHAP explanation of the issue of a machine learning model. \rev{The value is calculated using the sum over all four categories. The minimum value of $-4=1*4$ can be achieved, if all categories do not apply, the maximum value of $4=1*4$ is the result, if all categories apply. The overall value of $qs$ is then calculated using the average between all raters.}   
\end{itemize}
As independent variables, we use a one-hot encoding of the confusion of the prediction, such that we have

\begin{itemize}
\item $tp \in \{0, 1\}$: 1 if the prediction of the issue by the machine learning model is a true positive, 0 otherwise.
\item $tn \in \{0, 1\}$: 1 if the prediction of the issue by the machine learning model is a true negative, 0 otherwise.
\item $fp \in \{0, 1\}$: 1 if the prediction of the issue by the machine learning model is a false positive, 0 otherwise.
\item $fn \in \{0, 1\}$: 1 if the prediction of the issue by the machine learning model is a false negative, 0 otherwise.
\end{itemize}

Moreover, we consider the following confounding variables, i.e., variables that may be alternative explanations for our results.

\begin{itemize}
    \item $p_i \in \{0,1\}$ for $i=1,...,39$: A one-hot encoding of the project to which the issue belongs. 1 if the issue belongs to project $i$, 0 otherwise. This confounder considers if the quality of the explanations may also be explained by the projects, for which they are produced.
    \item $first_{lime} \in \{0,1\}$: 1 if the user has first seen the LIME explanation of the issue, 0 otherwise. This confounder measure if the order of the explanation presented to the rater may influences the rating. 
    \item $n \in [1,3092]$: number of issues seen by the rater. For the first presented issue is $n=1$, for the second issue is $n=2$, etc.. This confounder measure if the expectations of the rater changes over time and the rater has a different measurement for the quality of the explanation between the first issues and last issues. 
\end{itemize}

\subsection{Execution Plan}
The main task of the execution is to rate the LIME and SHAP explanations based on the quality and conduct a qualitative rating. Three of our co-authors will act as raters to assess all issues independently and measure the average of these three qualitative assessments that are the foundation of our variable $qs$.\footnote{We will add additional authors (graduate students, possibly PostDocs) from our still growing research group, when we execute this study.} 

The main task of the rater is to perform a qualitative rating on the LIME and SHAP explanation. We will show the explanation of LIME and SHAP of an issue to the rater side by side. At the bottom of both explanations, the user can rate the explanations based on the categories from Table \ref{tab:tags}. The order of the explanations is random and are measured with the confounding variables $first_{lime}$. After the user finished the rating, they can submit the data and a new issue is selected randomly. The order of the explanation in the user interface is randomized for each issue. 
One might note, that the execution plan could be switched to a AB/BA testing \cite{madeyski2018effect}. However, because we only have three raters, the population sizes of n=2 for AB and n=1 for BA would be extremely low leading to uncertain results. Consequently, we believe our choice with the order of the explanations as a randomized confounding variable $first_{lime}$ yields more trust worthy results.

\begin{figure}[h]
\includegraphics[width=0.5\textwidth]{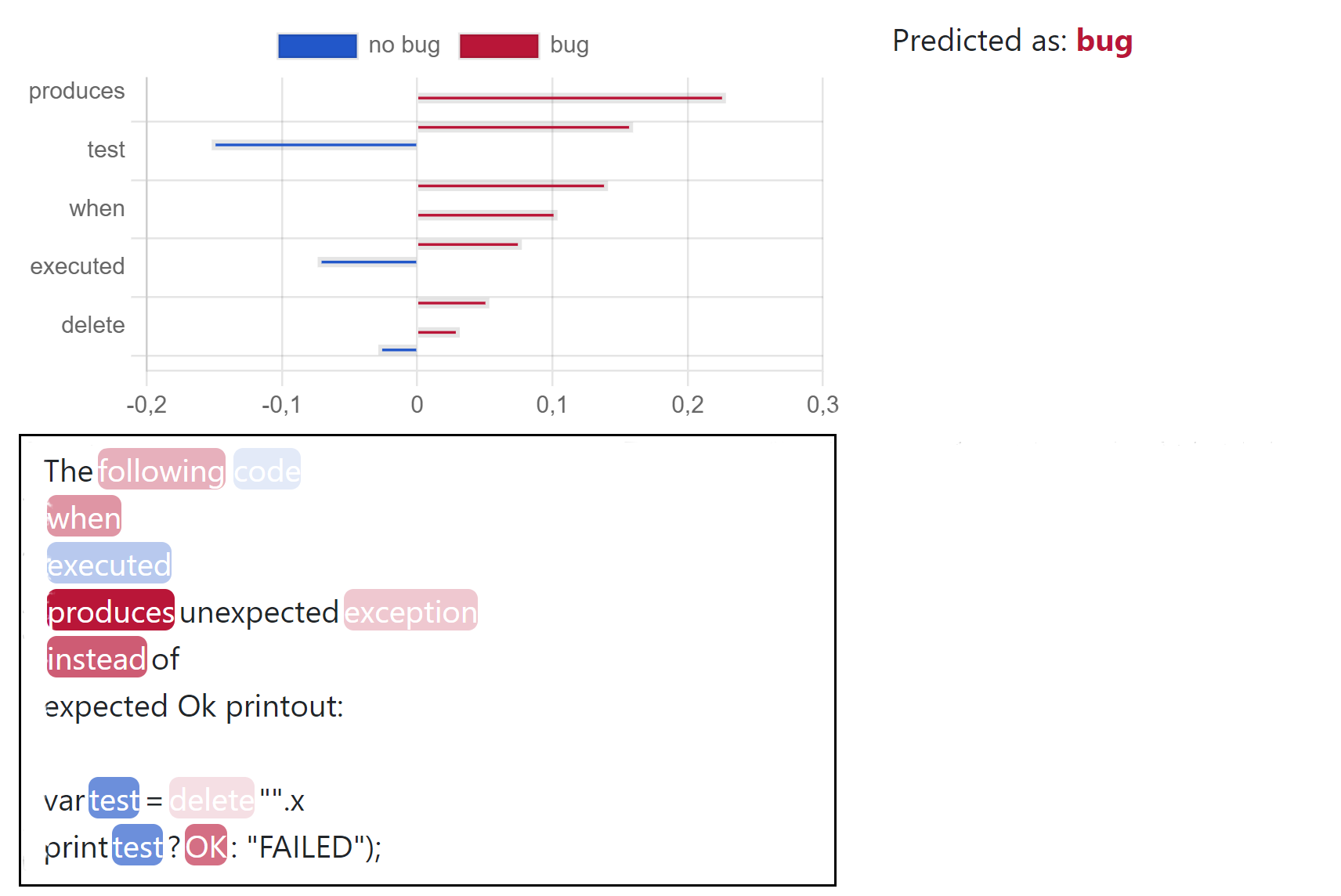}
\caption{Visualisation of the LIME or SHAP explanation of an issue description. We provide both a bar chart of the impact of different terms on the label, as well as an highlighting of these words within the text of the issue description. }
\label{figure:limevisualization}
\end{figure}

Figure \ref{figure:limevisualization} shows a screenshot of the labeling tool. It contains for each bug prediction approach the issue title and issue description. LIME and SHAP are able to calculate the relevant words from the title and description for the local prediction of the machine learning model \cite{ribeiro2016should}. We execute LIME with $10$ features. On the one hand, \rev{too} few features may result in an inadequate explanation \cite{zhang2019should}. \rev{On the other hand, since LIME already chooses the best local approximation for the amount of features, more features do not increase the user's level of information.} \rev{This parameter is not required by SHAP. The result of SHAP are group of words. In order to make it equivalent to the LIME explanations, we also choose the ten most pertinent word groupings. But frequently more than one word is included in each group of words.} We will use Deep SHAP with the optimization for the natural language models \footnote{\url{https://github.com/slundberg/shap\#natural-language-example-transformers}}. The optimization are adding coalitional rules to traditional Shapley values. This helps to explain large modern natural language models using few function evaluations. 

We will display the five-most-relevant words from the title and the ten-most-relevant words from the description for being a bug and a non-bug. The bug words are highlighted in red and the non-bug in blue. LIME and also SHAP computes a scores for each of the words \cite{ribeiro2016should}, reflecting the relevance of the word to the outcome. We adjust the color strength based on the score. The word with the highest score achieves an alpha value of 1. The other words are linearly adjusted. Additionally, a bar chart is displayed on top of the issue to support the explanation. 

We order the explanation of the predictions by their quality as assessed by the ratings for the categories displayed in Table \ref{tab:tags}. The rater have to decide for each category applies, is neutral or does not apply for each of the explanations individually. It is only allowed to submit the result for the complete issue at once. We only rate the quality of the explanation, not of the prediction itself. Thus, the true label of the issue is not displayed for the rater.

Each rater starts with a short tutorial that will explain the functionality of the toolkit to avoid wrong ordering of the explanations. Additionally, we will explain the definition of bug issues based on Herzig et al. \cite{herzig2013s}, which is used in the data to achieve a common knowledge among the raters.
After completion, the rater starts with the first issue (either LIME or SHAP). The issue is automatically selected randomly from the sampled issues from all projects that are not already completed. The rater is allowed to skip an issue; However, we will measure the position of the issue with the confounding variable $n$. Each issue that is selected according to the sampling strategy is shown to each rater of the study. Thus, each rater has to rate 3,092 issues for both explanation algorithm. 

\subsection{Analysis Plan}
To answer our research question and hypotheses, we will execute the following three different analysis phases on our collected data. Each of the phases answers one of our hypotheses.

The core of our analysis is a linear model, which we use to determine the relationship between our variables. We then analyse the coefficients of the linear models to evaluate our hypotheses. As input for the fitting of the linear model, we use all pairs of issues and predictions (see Section~\ref{sec:participantstags}). We measure the goodness-of-fit of the models with the $R^2$ coefficient. 

Additionally, we provide the distributions of the dependent variable $qs$ for each of the explanation models, both overall, as well as grouped by the predicted labels. To enable deeper insights into the qualitative ratings, we also provide the distributions of the separate quality categories of the explanations. These distributions are used to augment the analysis of the hypothesis and the resulting discussion of the results, to provide insights into the explanations beyond the pure statistical analysis. 

To gain insights into the validity of our data, we measure the agreement between raters. We report Fleiss' $\kappa$~\cite{Fleiss1971} to estimate the reliability of the qualitative ratings for each category listed in Table~\ref{tab:tags}, which is defined as
\begin{equation}
\kappa = \frac{\bar{P}-\bar{P}_e}{1-\bar{P}_e},
\end{equation}
where $\bar{P}$ is the mean agreement of the participants per issue and $\bar{P}_e$ is the sum of the squared proportions of the label assignments. We use the table from Landis and Koch~\cite{Landis1977} for the interpretation of $\kappa$ (see Table~\ref{tbl:kappa}).

\begin{table}
\centering
\begin{tabular}{cl}
$\rho$ & \textbf{Interpretation} \\
\hline \hline
$<0.1$ & No correlation \\
0.10 – 0.29 & weak correlation \\
0.30 – 0.49 & moderate correlation \\
0.50 – 1.00 & strong correlation \\
\end{tabular}
\caption{Interpretation of Spearman's $\rho$ according to Cohen~\cite{Cohen1988}.}
\label{tbl:rho}
\end{table}

\begin{table}
\centering
\begin{tabular}{cl}
$\kappa$ & \textbf{Interpretation} \\
\hline \hline
$<$0 & Poor agreement \\
0.01 – 0.20 & Slight agreement \\
0.21 – 0.40 & Fair agreement \\
0.41 – 0.60 & Moderate agreement \\
0.61 – 0.80 & Substantial agreement \\
0.81 – 1.00 & Almost perfect agreement \\
\end{tabular}
\caption{Interpretation of Fleiss' $\kappa$ according to Landis and Koch~\cite{Landis1977}.}
\label{tbl:kappa}
\end{table}
Second, we repeat this measurement for the Spearman's $\rho$ of $qs$ for each category of Table \ref{tab:tags} separately, which allows us to gain insights about differences in the four categories. Spearman's $\rho$ is defined as
\begin{equation}
\rho = \frac{cov(RS_1, RS_2)}{\sigma_{RS_1}\sigma_{RS_2}},
\end{equation}
where $RS_1, RS_2$ are the ranks of two raters, and $\sigma_{RS_1}, \sigma_{RS_2}$ the standard deviations of the ranks. We use the table from Cohen~\cite{Cohen1988} for the interpretation of $\rho$ (see Table~\ref{tbl:rho}).

\subsubsection{Hypotheses H1}
We use a linear model of $qs$ based on our independent and confounding variables, i.e., 
\begin{equation} \label{eq:h2regmodel}
\begin{aligned}
qs = b_0 + b_1\cdot tp +b_2\cdot tn + b_3\cdot fp + b_4\cdot fn + b_5\cdot first_{lime} \\ + b_6\cdot n + \sum_{i=1}^m b^p_i \cdot p_i,
\end{aligned}
\end{equation}
where $p_i$ are the one-hot encoded projects for our $m=39$ projects. \rev{We expect that the coefficient for $tp$ is greater than the coefficient for $tn$ as well as a significant difference between the subsets with a non-negligible effect size.}

\subsubsection{Hypotheses H2}
We use the linear model derived from Equation \ref{eq:h2regmodel} to analyse Hypotheses H2, i.e., we consider the magnitudes of the coefficients for $tp$ and $tn$. Additionally, we evaluate the statistical difference between the $qs$ on the subsets of issues that are true positives versus the subset of issues that are true negatives. Depending on the normality of the data, we will either use the t-test with Cohen's $d$ to measure the effect size, or the Mann-Whitney-U test with Cliff's $\delta$ as effect size. \rev{ We expect that $p_i$ are associated with a small coefficient and that there is no correlation between $p_i$ and $qs$.}

\subsubsection{Hypotheses H3} 
We use the average $qs$ of LIME and SHAP to analyse Hypotheses H3, i.e., we evaluate the statistical difference between the $qs$ of the subsets of LIME and SHAP explanations. Depending on the normality of the data, we will either use the paired t-test with Cohen's $d$ to measure the effect size, or the Wilcoxon signed-rank test with Cliff's $\delta$ as effect size. We expect, that $qs$ of the subset of SHAP explanation is significantly greater than of the subset of LIME explanation with a non-negligible effect size.

\section{Limitations}
\label{sec:limitation}
\rev{One of the limitations} of this study is that the relationship between explanation and prediction may be dependent on the machine learning model we consider. Other algorithms may have different relationships. For example, algorithms with an even better performance may further improve explanations. However, it is also possible that the quality of LIME and SHAP explanations are in general limited to some degree and, therefore, is at some point decoupled from additional improvements of the prediction performance. Furthermore, we cannot be sure if our observations regarding the relationship between model quality and the explanations generalized beyond the SE domain to other applications. 

Moreover, the overall stability of LIME explanations is a threat to our study. Visani et al. \cite{garreau2020looking} and Zhang et al. \cite{zhang2019should} found that LIME is not always stable. Based on that work, Visani et al. \cite{visani2020statistical} introduced different approaches to measure the stability of a LIME explanations. Another limitation in our study is that we label the data in our own. We mitigate that by selecting at least three different person from our research group and calculate the agreement between the raters to measure bias in the labeled data.

\section{Publication of generated data}

All materials we use are already publicly available in long-term archives. All generated data will be shared in a DOI citable long-term archive with an Apache 2.0 license or Creative Commons license. At least the following data will be published: The LIME and SHAP explanations and visual representation of all issues that are used in the study, including the qualitative rating of the explanation. This includes a pseudonym of the raters that conducted the rating. 

\section{Conclusion}

Within this study, we plan to shed light on the inner workings of the prediction of bug issues by manually analysing LIME and SHAP explanations. Through this study, we will increase our \rev{understanding} of the identification of bug issues, as well as the relationship between LIME, SHAP explanations and model quality. 

\bibliographystyle{ACM-Reference-Format}
\bibliography{main}

\end{document}